**Deep Learning Models in Speech Recognition: Measuring GPU Energy Consumption, Impact of Noise and Model Quantization for Edge Deployment**

Aditya Chakravarty, PhD
Texas A&M University

## Abstract

Recent transformer-based ASR models have achieved word-error rates (WER) below 4%, surpassing human annotator accuracy, yet they demand extensive server resources, contributing to significant carbon footprints. The traditional server-based architecture of ASR also presents privacy concerns, alongside reliability and latency issues due to network dependencies. In contrast, on-device (edge) ASR enhances privacy, boosts performance, and promotes sustainability by effectively balancing energy use and accuracy for specific applications. This study examines the effects of quantization, memory demands, and energy consumption on the performance of various ASR model inference on the NVIDIA Jetson Orin Nano. By analyzing WER and transcription speed across models using FP32, FP16, and INT8 quantization on clean and noisy datasets, we highlight the crucial trade-offs between accuracy, speeds, quantization, energy efficiency, and memory needs. We found that changing precision from fp32 to fp16 halves the energy consumption for audio transcription across different models, with minimal performance degradation. A larger model size and number of parameters neither guarantees better resilience to noise, nor predicts the energy consumption for a given transcription load. These, along with several other findings offer novel insights for optimizing ASR systems within energy- and memory-limited environments, crucial for the development of efficient on-device ASR solutions. The code and input data needed to reproduce the results in this article are open sourced are available on [https://github.com/zzadiues3338/ASR-energy-jetson].

## Introduction

Automatic Speech Recognition (ASR) transforms spoken language into written text. This technology has various applications such as dictation, accessibility features, hearable devices, voice-controlled assistants, and augmented/virtual reality environments. The field has seen significant advancements recently due to the adoption of deep learning methodologies for training and inferencing [Hinton et al., 2012]. Typically, cutting-edge ASR models operate on cloud-based frameworks where audio data from users is transmitted to servers for processing, with the results then sent back to the user. This setup, however, raises concerns about privacy and security of sensitive audio data and suffers from potential issues related to network dependency, such as latency and reliability.

Generally, the focus in machine learning is on maximizing accuracy to achieve top results. However, research in GreenAI [Schwartz et al., 2020] indicates that beyond a certain point, minor accuracy enhancements (measured in Word Error Rate, or WER) lead to a disproportionately high increase in energy consumption for training and inference. The environmental impact of these technologies could exacerbate the global energy crisis as the use of deep learning in AI applications becomes more widespread [Cao et al., 2020 and, Strubell et al., 2020]. For ASR specifically, a more balanced approach might involve selecting models based on their energy consumption, accuracy, and overall performance. For instance, in on-device dictation apps, occasional minor errors might be acceptable if the application allows users to correct them through a keyboard or other input methods. While a lower WER generally enhances user experience, prioritizing a slightly higher WER that conserves battery life could be a wiser choice, supplemented by other user interface solutions to meet application needs.

This paper explores these trade-offs in energy and accuracy for ASR processing at the edge, where it inherently safeguards user privacy and improves reliability by eliminating the need for internet connectivity. On-device ASR not only enhances data security and performance but also proves more energy-efficient. Unlike cloud processing that requires streaming audio to a server, on-device processing saves the energy costs associated with Wi-Fi/LTE connections, data transmission [Wu et al., 2019], and server computations, which are more energy-intensive due to the infrastructure of data centers.

Previous research has explored on-device ASR capabilities using models like on other platforms like the Raspberry Pi [Gondi et al., 2021] DeepSpeech [Hannun et al., 2014], and Wav2Letter [Pratap et al., 2019]. The goal of this study is to quantify the trade-offs between accuracy, energy, and performance for ASR using a typical transformer-based model on edge devices. To the best of the author's knowledge, this is the first study to evaluate ASR inference on edge GPU devices in terms of efficiency and performance. The NVIDIA Jetson Orin Nano was chosen for this project because it features a dedicated graphics processing unit (GPU), which is essential for efficiently processing deep learning tasks directly on the device. The Orin Nano strikes an optimal balance between power and compactness, making it ideally suited for edge AI applications. Its lightweight design allows for seamless integration into smaller devices, while still providing the computational capabilities necessary to handle complex AI loads without reliance on cloud computing resources. This makes the Jetson Orin Nano an excellent choice for developing energy-efficient, high-performance AI applications at the edge. The main contributions of this paper are outlined as follows:

- Benchmarking ASR Models on GPU: We evaluate several leading ASR models by analyzing their Word Error Rate (WER) and power consumption on GPU-enabled devices.
- Impact of Noise on WER: An assessment is conducted to understand how ambient noise in the dataset influences the accuracy (WER) of these models.
- Comprehensive Performance Evaluation: A detailed analysis is provided that considers the effects of model size, quantization type, and the impact of noise and model type on key performance metrics such as processing time, WER, and power usage.

- Optimal Configurations for Edge Deployment: Our findings aim to identify the most efficient and effective configurations for deploying ASR technologies in environments where power consumption is a critical concern.

The rest of the paper is structured as follows: In the background section we discuss high-level working of ASR and the models used in this study. In the experimental setup, we go through the steps for preparing the model and experimental setting up for model inferencing and energy consumption measurements. In the results section we cover the impact of quantization, noise in audio signal, and their interplay with model performance. Finally, we conclude with a summary and outlook.

**Automatic Speech Recognition**

ASR, or Automatic Speech Recognition, involves transforming audio signals into text. Fundamentally, ASR employs an acoustic model, a pronunciation model, and a language model to transcribe raw audio into text. With the advent of deep learning (DL) systems components of ASR have been enhanced using architectures like CNNs [Gu et al., 2018] and RNNs [Graves, 2012]. This led to the development of end-to-end and sequence-to-sequence models that streamline both the training and deployment of ASR systems. The most effective end-to-end systems utilize techniques such as connectionist temporal classification (CTC) [Graves et al., 2016], recurrent neural network transducers, and attention-based encoder-decoder frameworks [Bahdanau et al., 2017]. The transformer, initially developed for machine translation [Vaswani, et al., 2017], has been adapted for ASR to handle audio inputs rather than text [Baevski et al., 2020]. Optimized Attention architectures have recently been shown to improve the power efficiency of ASR models (Li et al., 2023).

ASR is demanding in terms of CPU/GPU compute resources and memory usage. Key considerations for implementing ASR on edge devices include CPU/GPU capacity, memory, storage, energy/battery life, and thermal management, especially for mobile or wearable technology. Challenges include fewer model updates compared to cloud-based systems, which allow for continuous enhancements and quick fixes, a lack of real-time data for ongoing model refinement, and insufficient metadata for debugging [Véstias et al. 2019]. Despite these issues, the advantages of on-device processing, such as enhanced privacy, reliability, and energy efficiency, outweigh these challenges and must be addressed by the edge AI community.
The tests performed for this study used the following models: whisperx (Bain et al., 2023), distil-whisper (Gandhi et al., 2023), speechT5 (Ao et al., 2022), wav2vec2 (Baevski et al., 2020), HuBERT (Hsu et al., 2021), WavLM (Chen et al., 2023). UniSpeech (Wang et al., 2021) and SpeechLM (Zhang et al., 2022). Table 1 provides a summary of the models used in this study.

**Table 1**: Summary of ASR models used in the study.

| Model name | Params, M | Size on disk, MB | Compute Type |
|---|---|---|---|
| whisperx-tiny | 39 | 151 | float32 |
| whisperx-tiny | 39 | 151 | float16 |
| whisperx-tiny | 39 | 151 | int8 |
| whisperx-small | 244 | 967 | float32 |
| whisperx-small | 244 | 967 | float16 |
| whisperx-small | 244 | 967 | int8 |
| whisperx-base | 74 | 290 | float32 |
| whisperx-base | 74 | 290 | float16 |
| whisperx-base | 74 | 290 | int8 |
| whisperx-medium | 769 | 1492 | float16 |
| whisperx-medium | 769 | 1492 | int8 |
| distil_whisper-large-v2 | 756 | 1510 | float16 |
| distil_whisper-medium | 394 | 790 | float16 |
| distil_whisper-small | 166 | 332 | float16 |
| Hubert-large | 315 | 1260 | float16 |
| wavLM-base | 94 | 378 | float16 |
| unispeech | 315 | 1270 | float16 |
| speecht5_asr | 154 | 606 | float16 |
| wav2vec2-large | 315 | 1270 | float16 |

**Data**

70 audio tracks in .flac format from a subdirectory from the LibriSpeech (Panayotov et al., 2015) open database (https://openslr.org/) were used as inputs for the text transcription. The audio frequency is 16 kHz, has very high signal quality i.e., high signal to noise ratio and audio duration varies between 3 seconds to 5 seconds.

**Noise addition to clean audio data**

In the experimental setup, each audio track was subjected to a noise corruption process to simulate real-world audio degradation. Gaussian white noise was synthetically generated with a zero mean ($\mu=0$) and unit standard deviation ($\sigma=1$). The noise was sampled to match the discrete length of the audio track's sample array, ensuring synchronous additive noise overlay. After its generation, the noise amplitude was attenuated by 10 decibels to temper its influence on the underlying signal. The attenuated noise was then algebraically added to the original audio track to produce a noisy version (Figure 2). This mixture was achieved using an overlay process that maintains the original audio track's temporal dynamics while introducing the stochastic characteristics of the white noise, thereby creating a composite signal that mirrors potential real-life audio signal corruption encountered in noisy environments.

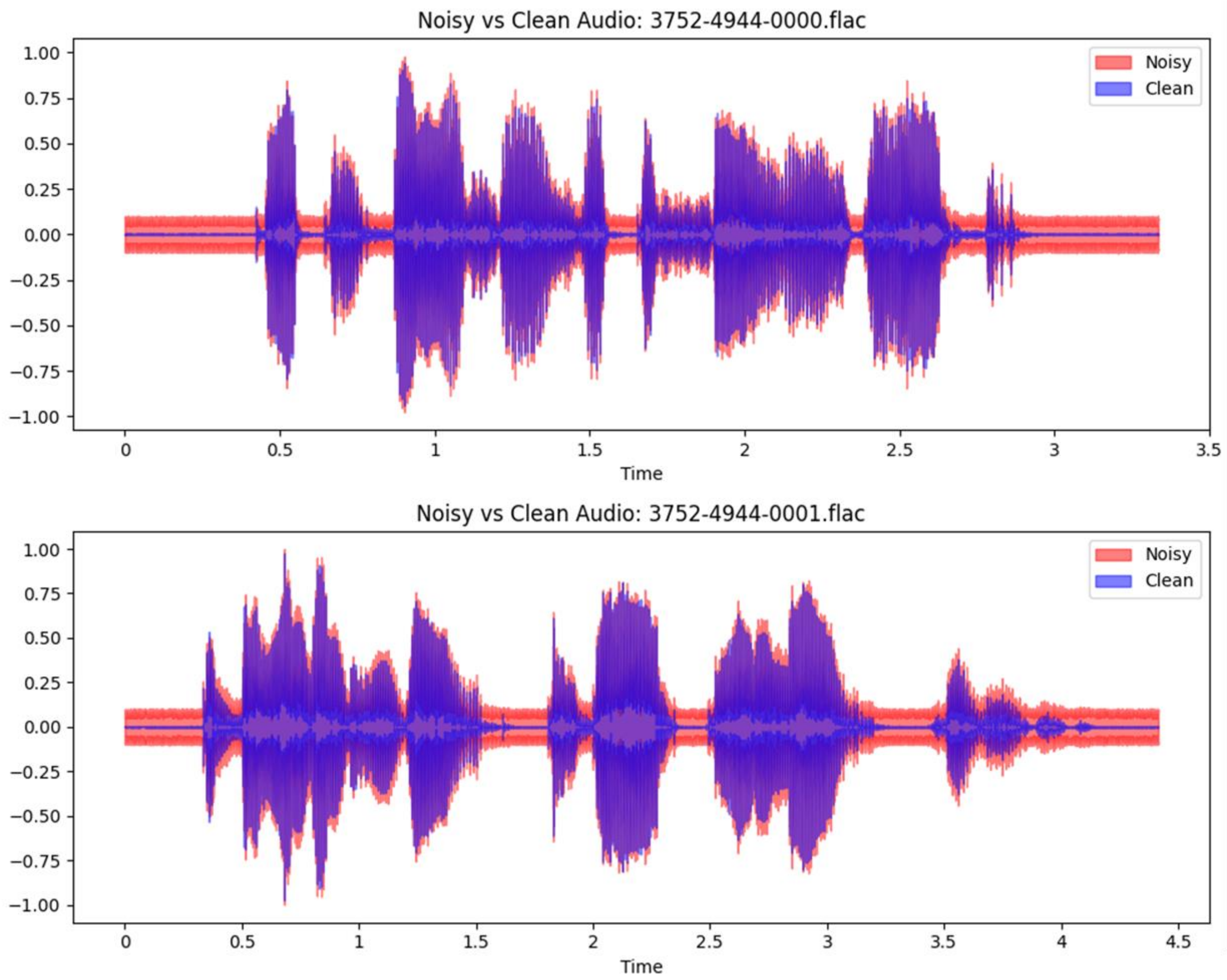

**Figure 2**: Two examples of the input audio waveforms from the Librispeech audio corpus. Input to the models is a .flac file and output is txt file containing the transcripted text. In the examples shown, the clean audio sample (blue) is superimposed on the same audio with added Gaussian white noise (red). Duration of input files lie between 3 to 5 seconds.

## GPU Device and Experimental Setup

The GPU device is interfaced through wireless keyboard and mouse, and the graphical user interface is done through the JetPack 5.1 OS that is installed on the Orin Nano (Figure 3). The deep learning code (Github link for the code provided) is run using Docker container due to overcome managing the complex library dependencies on the JetPack OS. Model inference is performed using a combination of the jetson-containers (https://github.com/dusty-nv/jetson-containers) and the Huggingface framework. Detailed information about power consumption was extracted using the 'tegrastats' utility, which is inbuilt in the OS. Establishing baseline values for RAM and power consumption is imperative to determine the precise allocation of resources for deep learning tasks.

Baseline measurements of RAM and power were ascertained by running tegrastats for one minute during a period of inactivity on the Jetson device. This procedure resulted in baseline figures of 515 mW for CPU and GPU power combined, along with a RAM usage of 2273 MB. This baseline load originates from the Linux GUI and system idle processes. These baseline values were then subtracted from the total power and RAM consumption recorded during deep learning tasks to calculate the specific resources allocated to those processes. Each run of inference generated a log file detailing memory usage, model performance, and energy consumption, and the aggregation of these log files formed the basis of the raw data collected for subsequent evaluation.

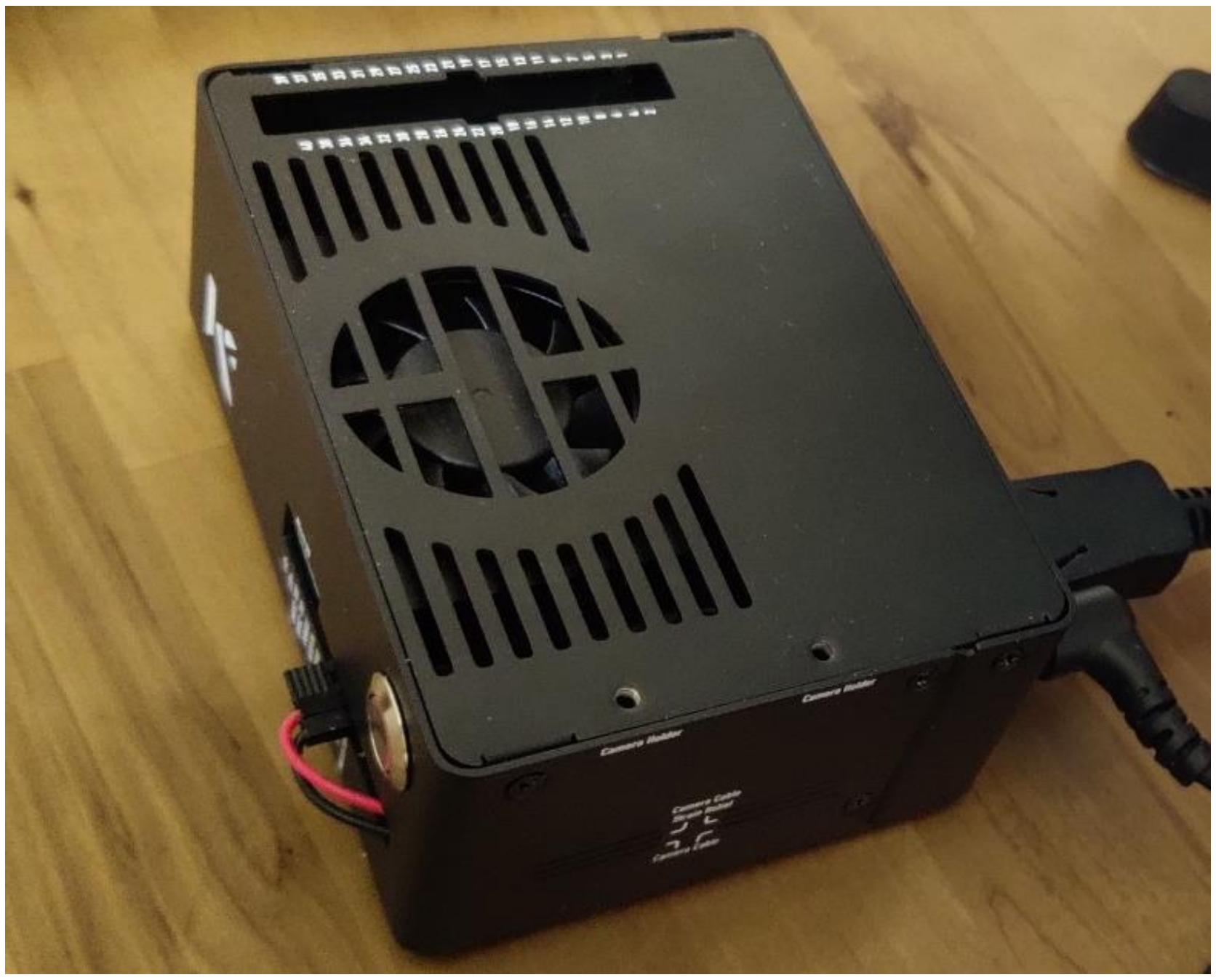

**Figure 3:** The NVIDIA Jetson Orin Nano board is housed inside a stainless-steel case. At the back are the power inlet, ethernet port, HDMI port and USB slots. The maximum power is 15 W, and the board is cooled by a fan assembly.

**Table 2**: NVIDIA Jetson Orin Nano technical specifications

| | |
|---|---|
| GPU | NVIDIA Ampere architecture with 1024 CUDA cores and 32 tensor cores |
| CPU | 6-core Arm® Cortex®-A78AE v8.2 64-bit CPU 1.5MB L2 + 4MB L3 |
| Memory | 8GB 128-bit LPDDR5 68GB/s |
| Power | 7W - 15W |
| Dimensions | 100mm x 79mm x 21mm |

The ASR model inference output is a .txt file, each corresponding to a single .flac audio input file. The output text and input text undergo a ‘normalization’ step wherein punctuation marks, special characters and any capitalization are removed. This step is done to ensure consistency between the characters in prediction and ground truth datasets. The metric used to evaluate transcription quality is word error rate. Word error rate (WER) is a metric that measures how accurately an automatic speech recognition (ASR) system performs. It's calculated by comparing the number of errors in the transcription text produced by an ASR system to a human transcription. The formula for calculating WER is:

$$WER = (S+D+N)/N$$

where S: number of substitutions, D: number of deletions, N: number of insertions. The second metric that is calculated is to gauge the inference time of models and is termed Real time transcription factor (RTF) - defined as the time taken of inference divided by the duration of the audio. For both metrics, a lower value implies better performance.

**Results and discussion**

**Table 3**: Energy use and inference time for fp16 precision, clean audio (duration: 328 seconds)

| Model Name | Param, M | Size, MB | Time, s | RAM, MB | Energy, J | WER |
|---|---|---|---|---|---|---|
| whisperx-tiny | 39 | 151 | 64.3 | 4028 | 67 | 8.92 |
| whisperx-small | 244 | 967 | 151.7 | 4450 | 515 | 5.01 |
| whisperx-base | 74 | 290 | 78.0 | 4029 | 155 | 7.4 |
| whisperx-med | 769 | 1492 | 343.7 | 5260 | 1471 | 3.7 |
| DW-large-v2 | 756 | 1510 | 154.4 | 4260 | 825 | 3.92 |
| DW-medium | 394 | 790 | 101.7 | 3417 | 468 | 4.03 |
| DW-small | 166 | 332 | 62.2 | 2933 | 251 | 3.37 |
| Hubert-large | 315 | 1260 | 30.5 | 4131 | 47 | 2.39 |
| wavLM-base | 94 | 378 | 21.9 | 3397 | 25 | 12.62 |
| unispeech | 315 | 1270 | 27.4 | 4123 | 45 | 35.8 |
| speecht5_asr | 154 | 606 | 124.9 | 3616 | 168 | 11.86 |
| wav2vec2-lg | 315 | 1270 | 26.0 | 3995 | 37 | 4.13 |

**Impact of noise audio on ASR Model Performance**

The performance of various Automatic Speech Recognition (ASR) models in handling noise reveals a complex relationship between model architecture, memory usage, parameter count, and energy efficiency (Table 3). The Word Error Rate (WER) under noisy conditions acts as a pivotal indicator of model robustness, with its increase from clean to noisy conditions highlighting the susceptibility of each model to auditory interference.

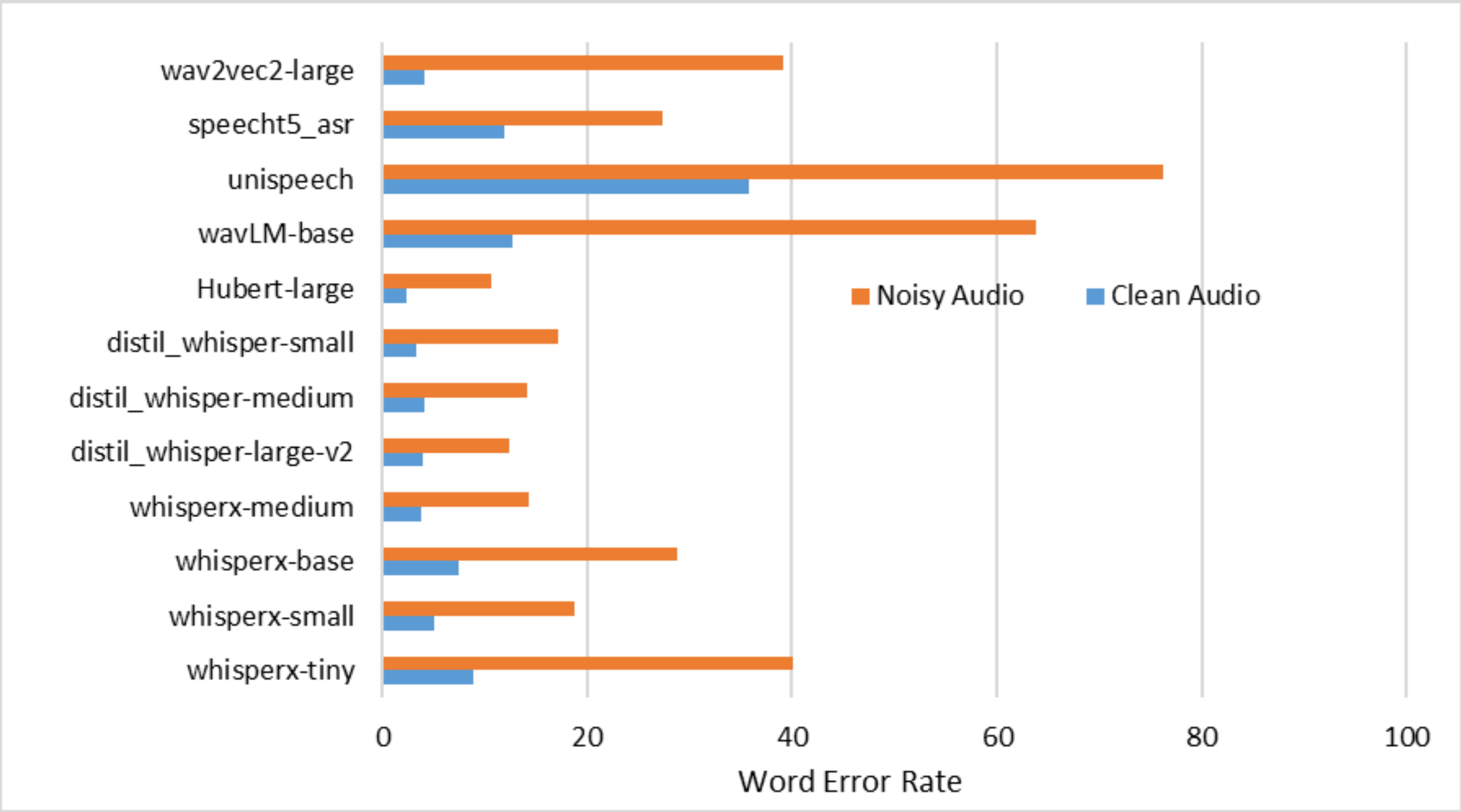

**Figure 4:** Impact of noisy audio input on the word error rate for tested models.

Larger models, such as whisperx-medium with 769 million parameters, and distil-whisper-large-v2 with 756 million parameters, demonstrate lower WERs in noisy environments (14.25 and 12.3, respectively), indicating that a higher number of parameters can potentially confer noise resilience (Figure 4). However, this comes at the cost of increased memory usage and energy consumption, as seen with whisperx-medium which uses 5260 MB of RAM and 1471 J of energy, suggesting a trade-off between noise robustness and computational resources. In contrast, HuBERT-large, despite having a moderate parameter count of 315 million, exhibits exceptional noise resilience (WER increasing from 2.39 to 10.66) with a relatively lower energy requirement (47 J). This suggests that parameter count alone does not dictate noise resilience and that model architecture plays a significant role.

On the other end of the spectrum, models with fewer parameters such as Distil-whisper-small (166 million parameters) and wav2vec2-large (315 million parameters) show a notable increase in WER when dealing with noise, indicating less robustness to acoustic disturbances. The Distil-whisper-small model, for instance, has a WER increase from 3.37 to 17.08, and while it consumes less RAM (2933 MB) and energy (251 J), its effectiveness in a noisy environment is compromised (Figure 4). It's noteworthy that memory usage does not consistently predict noise resilience, as wavLM-base with 3397 MB RAM usage underperforms in noisy conditions (WER of 63.87) compared to other models with similar or even higher memory demands.

In summary, the data reflects that while there is a tendency for models with larger parameter counts and higher energy consumption to maintain lower WERs in noisy conditions, this is not a universal rule. Models like HuBERT-large defy this trend by offering a balance between noise resilience, energy efficiency, and RAM usage. These insights underscore the necessity for a holistic approach to model selection for edge GPU deployment, considering not only the acoustic performance in diverse environments but also the operational constraints of power and memory.

**Impact of model quantization on energy consumption**

The influence of quantization on energy consumption in ASR models is markedly apparent when comparing the whisperx series across different data types (Figure 5). Quantization refers to the process of constraining an input from a large set to output in a smaller set, typically in the context of reducing the precision of the model weights [Gholami et al., 2021]. In the case of whisperx-tiny with 39 million parameters, quantizing from float32 to float16 reduced energy consumption by ~50 %, from 126 to 67 joules, a significant reduction that highlights the potential of lower-precision computations in energy savings. The whisperx medium model returned out of memory error in the fp32 precision mode. Further quantization to int8 only marginally increased energy use to 96 joules, still below that of the float32 implementation, suggesting a beneficial trade-off between energy efficiency and precision reduction.

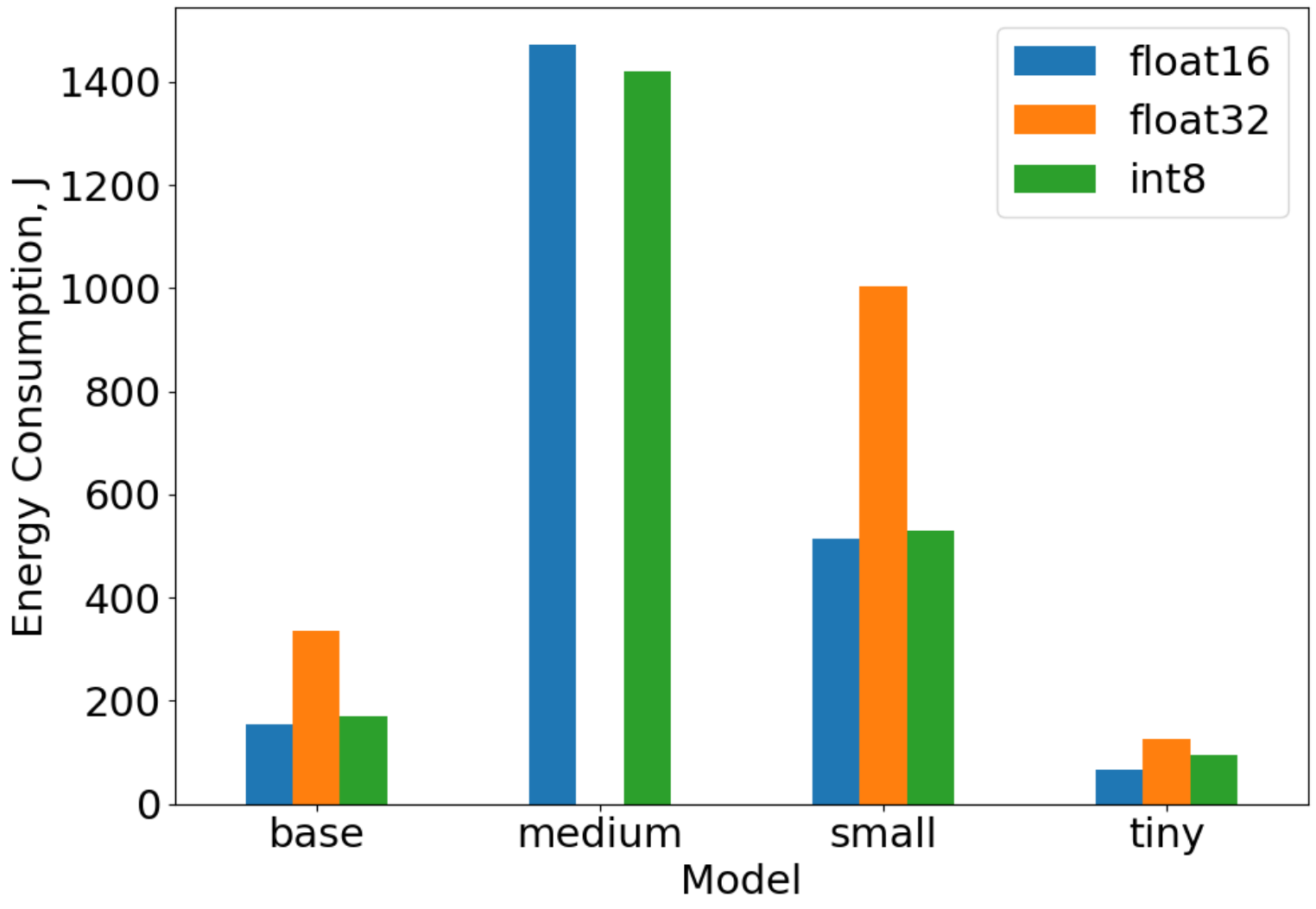

**Figure 5:** Impact of quantization on the energy consumption of whisperx models.

For whisperx-small and whisperx-base models with 244 and 74 million parameters respectively, we observe a similar pattern. The whisperx-small model saw a near halving in energy usage when quantized from float32 (1003 joules) to float16 (515 joules), with only a slight increase to 530 joules for the int8 version. Whisperx-base model's energy consumption also dropped from 336 joules in float32 to 155 joules in float16, and a minimal uptick to 169 joules for int8.

The whisperx-medium model, having the highest parameter count of 769 million, demonstrates the impact of quantization on larger models. With quantization from float16 to int8, energy consumption decreased from 1471 joules to 1420 joules. Although this reduction is less pronounced than in models with fewer parameters, it underscores the significance of quantization in optimizing energy efficiency across various model sizes. The consistent pattern across all models suggests that adopting lower precision data types through quantization can offer substantial energy savings, which is crucial for energy-constrained environments such as edge computing (Table 4). Notably, the reduction in energy consumption does not strictly correlate with a decrease in model size on disk, indicating that energy savings from quantization stem from reduced computational complexity rather than merely storage size. These findings are integral to the deployment strategies of ASR models, especially in applications where energy efficiency is paramount.

**Table 4:** WER, Energy consumption and RAM usage for whisperx quantized models

| Model | Precision | Time, s | Mem MB | Energy J | WER | WER-noisy |
|---|---|---|---|---|---|---|
| whisperx-tiny | float32 | 72.3 | 4273 | 126 | 8.92 | 40.04 |
| whisperx-tiny | float16 | 64.3 | 4028 | 67 | 8.92 | 40.04 |
| whisperx-tiny | int8 | 60.5 | 3944 | 96 | 8.92 | 41.57 |
| whisperx-small | float32 | 218.4 | 4892 | 1003 | 5.01 | 18.61 |
| whisperx-small | float16 | 151.7 | 4450 | 515 | 5.01 | 18.82 |
| whisperx-small | int8 | 155.3 | 4078 | 530 | 5.22 | 17.85 |
| whisperx-base | float32 | 97.8 | 4296 | 336 | 7.4 | 28.94 |
| whisperx-base | float16 | 78.0 | 4029 | 155 | 7.4 | 28.84 |
| whisperx-base | int8 | 80.1 | 3908 | 169 | 7.83 | 30.69 |
| whisperx-medium | float16 | 343.7 | 5260 | 1471 | 3.7 | 14.25 |
| whisperx-medium | int8 | 339.0 | 4651 | 1420 | 3.92 | 13.82 |

### Assessing impact of model quantization on WER in clean and noisy audio

Quantization's impact on WER degradation as audio transitions from clean to noisy is nuanced across the models tested (Figure 6). Notably, whisperx-tiny maintains identical WER in clean audio across float32 and float16 quantization but exhibits a slight increase with int8, hinting at a potential threshold in precision reduction that affects noise robustness.

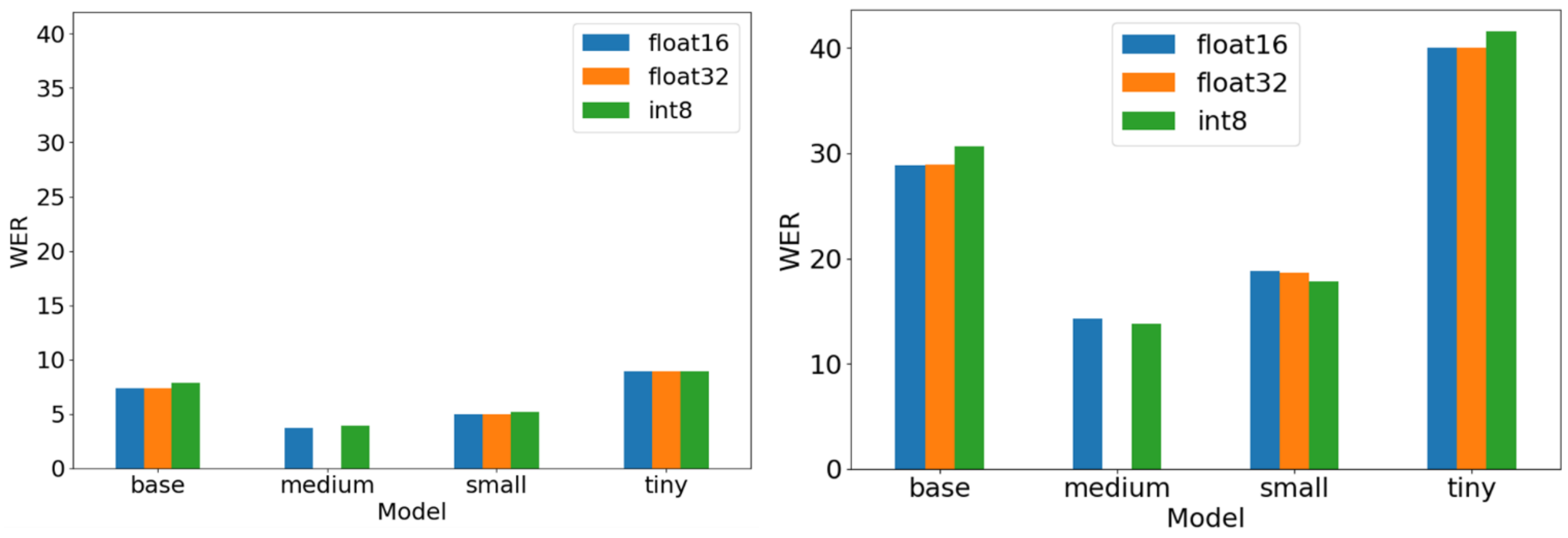

**Figure 6:** Impact of model quantization on word error rate (WER) for clean (left) and noisy audio (right) in whisperx models.

In larger models, such as whisperx-small and whisperx-base, the trend is less straightforward. For example, whisperx-small shows an unexpected dip in WER with int8 quantization in noisy audio, which may suggest a model-specific interaction between quantization and noise handling capabilities. The whisperx medium model returned out of memory error in the fp32 precision mode. Whisperx-medium presents a minimal increase in WER when moving from clean to noisy audio with int8 compared to float16, implying that the effect of quantization on noise performance may become less pronounced with increased model complexity. The slight uptick in WER with more aggressive quantization does not appear to be strictly tied to either disk size or memory usage, suggesting that the inherent noise processing ability of each model's architecture plays a more significant role in WER resilience amidst quantization levels. Considering the combined results of power consumption and WER, it's evident that using fp16 over fp32 precision leads to significant drop in energy consumption with negligible performance degradation.

**Impact of model quantization on model inference speed and memory usage**

The Real-Time Factor (RTF) reflects the speed of ASR models during transcription, with quantization type emerging as a significant factor influencing this metric (Figure 7, left). For whisperx-tiny, quantization from float32 to int8 results in a lower RTF, suggesting faster processing times and improved efficiency. This reduction in RTF is also observed in whisperx-small when moving from float32 to float16, although the switch to int8 does not yield further RTF improvement, indicating a possible optimal point of computational balance between float16 and int8 for this model. The whisperx medium model returned out of memory error in the fp32 precision mode. The whisperx-base shows a minimal difference in RTF between float16 and int8, implying a negligible impact of aggressive quantization on real-time performance for this model size. Moreover, the whisperx-medium model maintains a relatively high RTF even after quantization, which might be attributed to its larger size and higher RAM usage.

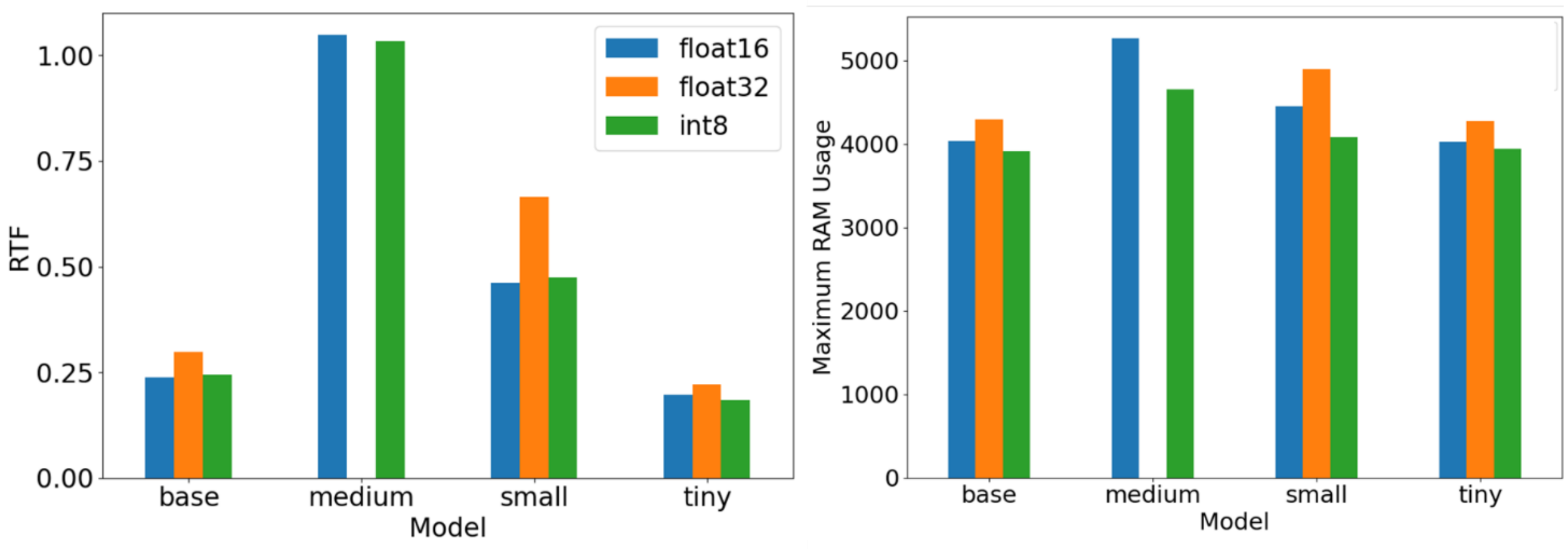

**Figure 7:** Impact of quantization on real time transcription factor, RTF (left) and the RAM usage (right) for whisperx models.

Quantization leads to a consistent reduction in memory requirements (Figure 7, right). For the whisperx-tiny model, quantization to float16 and int8 decreases memory usage from 4273 MB to 4028 MB and 3944 MB, respectively, demonstrating how lower precision formats can conserve memory resources. This pattern is consistent across the models-these results illustrate that memory usage declines with model quantization without necessarily affecting the size on disk proportionally.

**Summary and conclusion**

The interplay between Word Error Rate (WER), energy consumption, inference time, and memory usage present a multidimensional perspective on the performance of speech recognition models (Figure 8). Lower WER indicates better performance in clean audio, but this often correlates with higher energy usage and longer inference times, as seen with the whisperx-medium model. This model, while yielding a low WER of 3.7, consumes the most energy and has the longest inference time, emphasizing a trade-off between accuracy and efficiency.

Conversely, models like HuBERT-large achieve the lowest WER of 2.39 with much less energy and in a shorter time, showing that high efficiency and strong performance are achievable together. Memory usage does not show a direct correlation with WER, as models with both high and low memory footprints can achieve low WERs. The challenge lies in balancing these aspects; models must be optimized not just for accuracy but for operational efficiency, especially for applications where resources are constrained. This balance is crucial for the practical deployment of ASR models in real-world edge devices. The robustness of speech recognition models in the presence of noise significantly affects their practical utility, as indicated by the Noisy WER values (Figure 9).

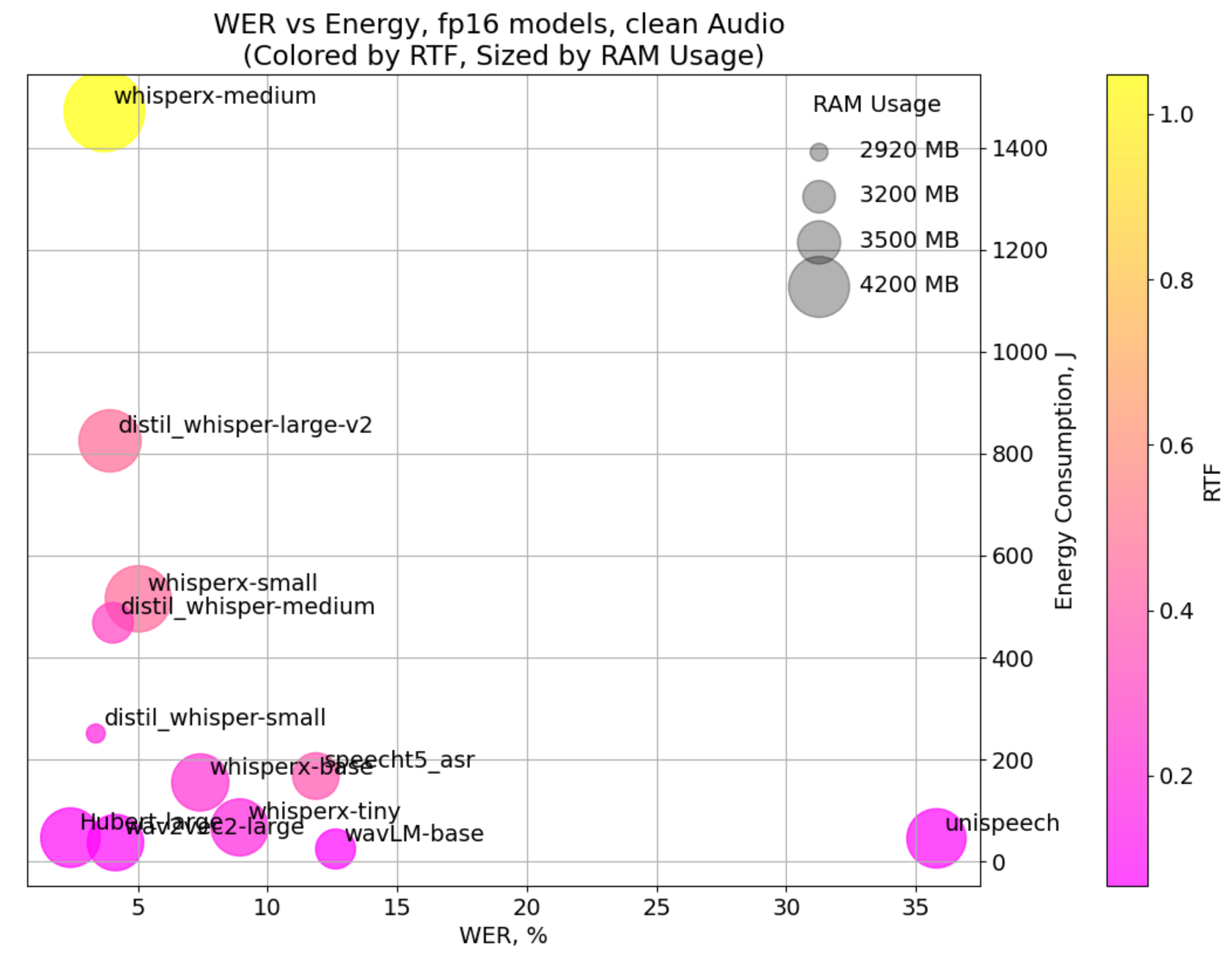

**Figure 8:** Scatterplot of word error rate and energy consumption for clean audio.

The whisperx-medium model shows remarkable resilience with its WER increasing to 14.25 in noisy environments, yet it demands substantial energy and has a high inference time, which could be a limitation for real-time applications. On the other hand, models like HuBERT-large not only maintain a low WER in clean conditions but also exhibit strong noise resistance, with a WER of 10.66 and lower energy consumption, positioning it as an optimal choice for efficiency and reliability.

However, models such as wavLM-base and unispeech struggle with noise, showing a significant spike in WER, despite lower energy usage and quicker inference times. This suggests that while some models are energy and time efficient, their performance under noisy conditions might render them unsuitable for environments where noise is unavoidable.

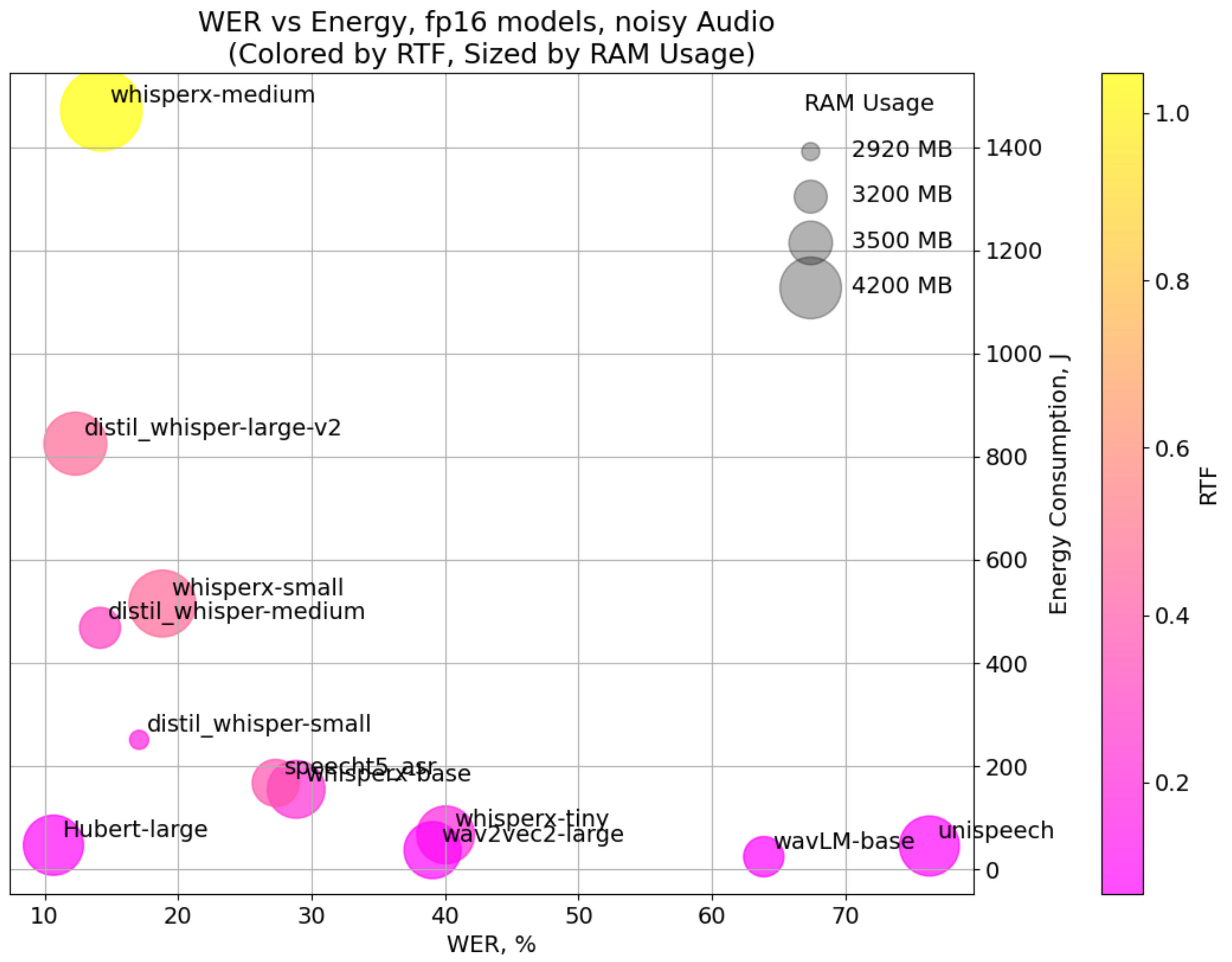

**Figure 9:** Scatterplot of word error rate and energy consumption for noisy audio.

Memory usage, once again, does not predict noise resilience, as models with both higher and lower memory usage show varying levels of noise WER. Considering the combined results of power consumption and WER, it's evident that using fp16 over fp32 precision leads to significant drop in energy consumption with negligible performance degradation.

This comparative analysis underscores the delicate balance between accuracy, energy consumption, inference speed, and memory usage in the deployment of speech recognition models on edge GPUs, especially under noisy conditions. The observations show that no single model excels in all areas (although, the HuBERT-large model comes quite close). Thus, the choice of model for deployment will depend heavily on the specific constraints and requirements of the use case. The ideal model would offer a blend of low WER in noisy conditions, reasonable energy consumption, and fast inference times.